# Interweaving Polar Charge Orders in a Layered Metallic Super-atomic Crystal

Shuya Xing[1,+], Linlu Wu[1,+], Zilu Wang[1,+], Xu Chen[2,+], Haining Liu[3,4], Shuo Han[1], Le Lei[1], Linwei Zhou[1], Qi Zheng[2,4], Li Huang[2,4], Xiao Lin[4], Shanshan Chen[1], Liming Xie[3,4], Xiaolong Chen[2,4,5] Hong-Jun Gao[2,4], Zhihai Cheng[1,*], Jiangang Guo[2,5,*], Shancai Wang[1,*], and Wei Ji[1,*]

[1]*Beijing Key Laboratory of Optoelectronic Functional Materials & Micro-nano Devices, Department of Physics, Renmin University of China, Beijing 100872, China*
[2]*Beijing National Laboratory for Condensed Matter Physics, Institute of Physics, Chinese Academy of Sciences, P.O. Box 603, Beijing 100190, China.*
[3]*CAS Key Laboratory of Standardization and Measurement for Nanotechnology, CAS Centre for Excellence in Nanoscience, National Centre for Nanoscience and Technology, Beijing 100190, China*
[4]*University of Chinese Academy of Sciences, Beijing 100039, China*
[5]*Songshan Lake Materials Laboratory, Dongguan, Guangdong 523808, China*

Electronic properties of super-atomic crystals have not been sufficiently explored due to the versatility of their building units; moreover, their inter-unit couplings are even poorly understood. Here, we present a joint experiment-theory investigation of a rational-designed layered super-atomic crystal of $Au_6Te_{12}Se_8$ cubes, stacked by non-covalent inter-cube quasi-bonds. We found a sequential-emerged anisotropic triple-cube charge-density-wave (tc-CDW) and polarized metallic states below 120 K, as revealed via scanning tunneling microscopy/spectroscopy, angle-resolved photoemission spectroscopy, transport measurement, Raman spectra, and density functional theory. The polarized states are locked in an anti-parallel configuration, which is required for maintaining the inversion symmetry of the center-cube in the tc-CDW. The anti-polar metallic states are thus interweaved by the charge-density-wave and the polarized metallic states, and primarily ascribed to electronic effects via theoretical calculations. This work not only demonstrates a microscopic picture of the interweaved CDW and polarized charge orders in the super-atomic crystal of ATS, but also sheds light on expanding the existing category of quantum materials to non-covalent solids.

[+] These authors contributed equally: Shuya Xing, Linlu Wu, Zilu Wang, Xu Chen.
* Email: zhihaicheng@ruc.edu.cn    jgguo@iphy.ac.cn    scw@ruc.edu.cn    wji@ruc.edu.cn

# I. INTRODUCTION

The pioneering synthesis of emergent crystals, e.g. iron-based superconductors (La[$O_{1-x}F_x$]FeAs)[1], perovskite solar cells ($CH_3NH_3PbI_3$[2, 3] and $CH_3NH_3PbBr_3$[4]), and two-dimensional magnetism ($CrI_3$)[5-7], has always attracted significant attention among researchers. In particular, these crystals have been useful in discovering novel physical phenomena in materials science, exploring intriguing properties, and unravelling unexplored principles. The subsequent optimization of these crystals, e.g. La[$O_{1-x}F_x$]FeAs, by altering one type of atoms with another was limited due to the fact that only 82 stable and non-radioactive elements could be used, and each of them has its own bonding characteristics which are largely different from each other. Thus, substituting a new type of atom into a given material typically introduces different structures with widely varying properties [8-10]. This, however, poses an overwhelming challenge to rationally designing functional materials [11, 12].

Stacking two-dimensional (2D) monolayers through noncovalent van der Waals (vdW) interactions between layers offers new means to overcoming the aforementioned challenge. In fact, vdW attractions in the stacked 2D materials usually introduce interlayer wavefunction overlaps with subsequent electronic hybridization. This is known as the covalent-like quasi-bonding (CLQB) because the overlapped wavefunctions yield a covalent-like charge redistribution characteristic at the interlayer region [13-16]. Exploiting CLQB is an interesting approach to introducing substantial changes to bandgaps [13, 14], optical transitions [15, 16], topological properties [17], magnetism [18, 19], electrical polarization [20], and superconductivity [10, 21], which are sensitively dependent on the configurations of involved 2D monolayers. This tunability is usually achieved through stacking [22], twisting [23], bending [24], or compressing [25] monolayers, which allows for a much higher flexibility and larger potential for tuning novel functionalities than the traditional approach of bulk crystals via covalent chemical bonding.

Although tremendous success has been achieved in vdW stacking of 2D-monolayers for building novel structures and systems, the noncovalent interaction prevails only in the stacking direction, while covalent bonding still governs the in-plane position and the type of atoms in each monolayer. Therefore, new strategies have been explored to introduce in-plane noncovalent bonding, namely, using super-atomic clusters instead of atoms, as building blocks to construct layered materials. Atomic clusters are also known as superatoms [12, 26-30]. These, in principle, have countless species available for materials design [12, 29, 30]. Here, we do not strictly require the electronic structures of a super-atom showing atomic-like orbital feature [31],such as those in the shells of S, P, D and others; in other words, we generalized the conventional concept of superatoms. Super-atomic crystals could be assembled through various linkage types that usually result in weak inter-superatom electronic interactions [12, 26-30]. As an emergent linkage, π-π CLQB helps in the formation of free-electron-like bands in pentacene [32] and $C_{60}$ monolayers [33, 34] under finely tuned compression. Nevertheless, electrons in these monolayers were well described in a single-particle picture, with the exception of strong electron–electron or electron–quasiparticle interactions that form the basis of many intriguing and complicated physical phenomena, such as superconductivity [35-37] and charge density wave (CDW) [37] observed in quantum materials.

Here, we show experimental evidence for the interweaved charge orders in a rationally designed (see Supplementary Note 1 for details of the design) layered material of cubically shaped $Au_6Te_{12}Se_8$ (ATS) super-atomic crystal (Fig.1), which offers up-to 12 Te…Te CLQBs [38, 39]. The Te…Te quasi-bonding in this case is considered a highly interesting noncovalent interaction form because of the reduced energy cost, high tunability, and strong electronic coupling, as illustrated in layered Te allotropes [40, 41]. The electronic structures and electrical properties of ATS were revealed using low-temperature scanning tunneling microscope (LT-STM), angle-resolved photoemission spectroscopy (ARPES) together with Raman and transport measurements, and density functional theory (DFT) calculations. We found two charge orders, namely a triple-cube-width stripe period along the *a*-axis and a spontaneous electric polarization with interlocked anti-polar directions along the *b*-axis, which were ascribed to strong electronic interactions. The electronically coupled super-atom layers belong to a category of layered materials that exhibit manipulatable novel properties in tremendous layered structures with high and precise tunability (see Fig. S1 for explanations).

## II. RESULTS

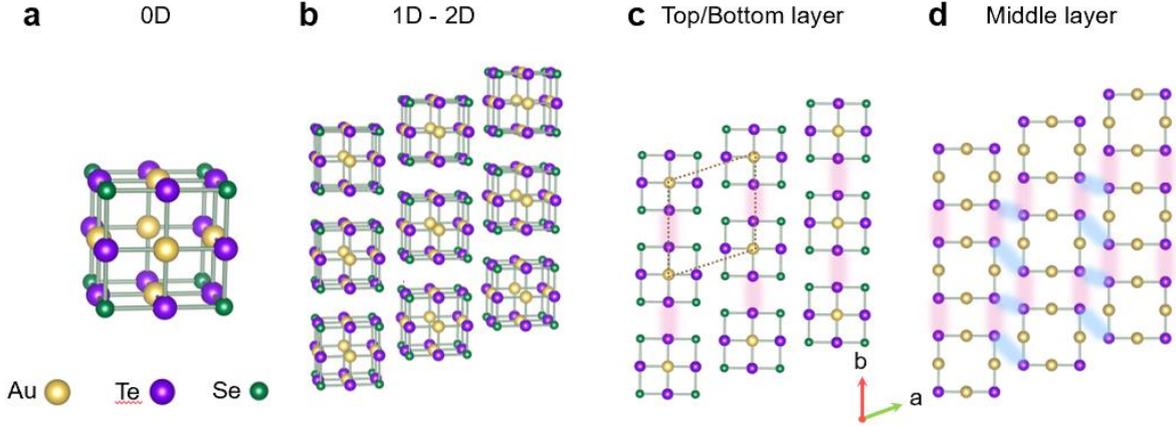

FIG. 1. Atomic structures and inter-cube interactions of the Au$_6$Te$_{12}$Se$_8$ (ATS) super-atomic crystal. (a) An individual ATS cube structure, regarded as a superatom (0D). (b) Atomic structure of an ordered ATS monolayer cleaved from an ATS bulk crystal, in which ATS cubes form one-dimensional (1D) chains along the *b*-axis. Top views for the slabs of the top/bottom (c) and middle (d) sublayers of the ATS monolayer, in which red and blue shadowed lines illustrate intra- (red) and inter-chain (blue) inter-cube Te…Te interactions, respectively. Yellow, purple, and green balls represent Au, Te, and Se atoms, respectively.

### A. Super-atomic crystal and CLQBs

An ATS superatom is composed of orderly arranged Au, Te, and Se atoms in a cube-like high-symmetric geometry, as shown in Fig. 1(a). The three types of atoms are, respectively, located at the faces, edges, and corners of the cubic structure. The lattice of a three-dimensional ATS super-atomic crystal may reduce its symmetry, producing *P*2 (*P*211) symmetry on a cleaving surface in the *ab*-plane (Fig. S2). In this plane, ATS superatom cubes form one-dimensional (1D) chains along the *b*-axis (the chain direction was originally defined as the *b*-axis in the bulk crystal), and they slightly slide across chains [Fig. 1(b) and Fig. S3]. In contrast to atoms (e.g., Te), the ATS superatoms have a higher inter-bonding (noncovalent) flexibility, which facilitates the formation of complicated and highly tunable 2D Te...Te quasi-bonding networks. In particular, there are four Te…Te noncovalent quasi-bonds [red shadowed lines in Fig. 1(c) and 1(d)]: one each in the top and bottom sublayers [Fig. 1(c)] and two in the middle sublayer [Fig. 1(d)]. These participate in the formation of ATS chains through noncovalent interactions. Two additional Te…Te bonds [blue shadow lines in Fig. 1(d)] were found only in the middle sublayer, suggesting strong anisotropy of inter-cube in-plane interactions within the ATS layer surface (Fig. S4). Figure S5 shows energy levels and $|\psi|^2$ of frontiers molecular orbitals (MOs) of an isolated ATS cluster. The highest occupied molecular orbital (HOMO), the lowest unoccupied molecular orbital (LUMO, doublet) and LUMO+1 (triplet) are separated from other molecular orbitals by at least 1.0 eV. These three MOs are primarily comprised of pronounced and extended out-of-plane Te-*p*, less significant Se-*p* and localized in-plane Au-*d* orbitals. The out-of-plane Te-*p* components of these three MOs further result in the strongest Te…Te inter-cluster interaction for those bands near the Fermi level of the ATS super-atomic crystal [Fig. S6&S7], while Te…Se and Te…Au interactions are substantially weaker.–

### B. Two sequentially emergent charge orders

Figure 2(a) shows the measured temperature-dependent resistivity and its first-order derivative (differential resistance) for an ATS flake sample. A nearly linear relation was observed for higher temperatures and the differential resistance begins to drop at ~ 90 K, with the slope becoming much steeper below 80 K. Such variation in differential resistance suggests the emergence of a likely charge order [42] at 90 K. Further decreasing the sample temperature results in a superconducting transition characteristic appearing at ~2.8 K (Fig. 2a and S8) [38]. However, the 90 K charge order transition was not reflected in our temperature-dependent Raman shift (TDRS) measurements [Fig. 2(b)], in which three (RP1/3/4) of the five TDRS peaks under our consideration (RP1-5) suggest a transition at ~120 K (see Fig. S9 for more details)." Interestingly, the two macroscopic measurements reveal different transition temperatures, stimulating interest into uncovering the origins of this discrepancy.

Figure 2(c–f) shows a series of temperature-dependent STM images and their corresponding FFT patterns (in the insets), which were acquired at 80, 100, 120 and 150 K, respectively. The results explicate two charge orders. In particular, the image acquired at 150 K [Fig. 2(f)] exhibits a chain-like appearance, which indicates no charge modulation and confirms the expected anisotropic inter-cube interactions. A triple-chain-width, stripe-like charge modulation appears in the a-axis at ~120 K [Fig. 2(e)] and becomes clearer at ~100 K [Fig. 2(d)]. An additional order emerges along the *b*-axis when the temperature decreases to ~80 K, which is showed in greater detail in an enlarged STM topographic image acquired at 9 K [Fig. 2(g)]. Surface lattice constants measured via STM ($a$ = 9.08 Å and $b$ = 9.32 Å) are found to be very close to those obtained via DFT calculations ($a = 9.02$ Å and $b = 9.28$ Å). The STM image shows multiple chain-like features parallel to the *b*-axis. Its FFT pattern confirms the stripe width of three ATS chains, which is in accordance with the observed characteristic peak ($q^* = a^*/3$) on the $a^*$ axis.

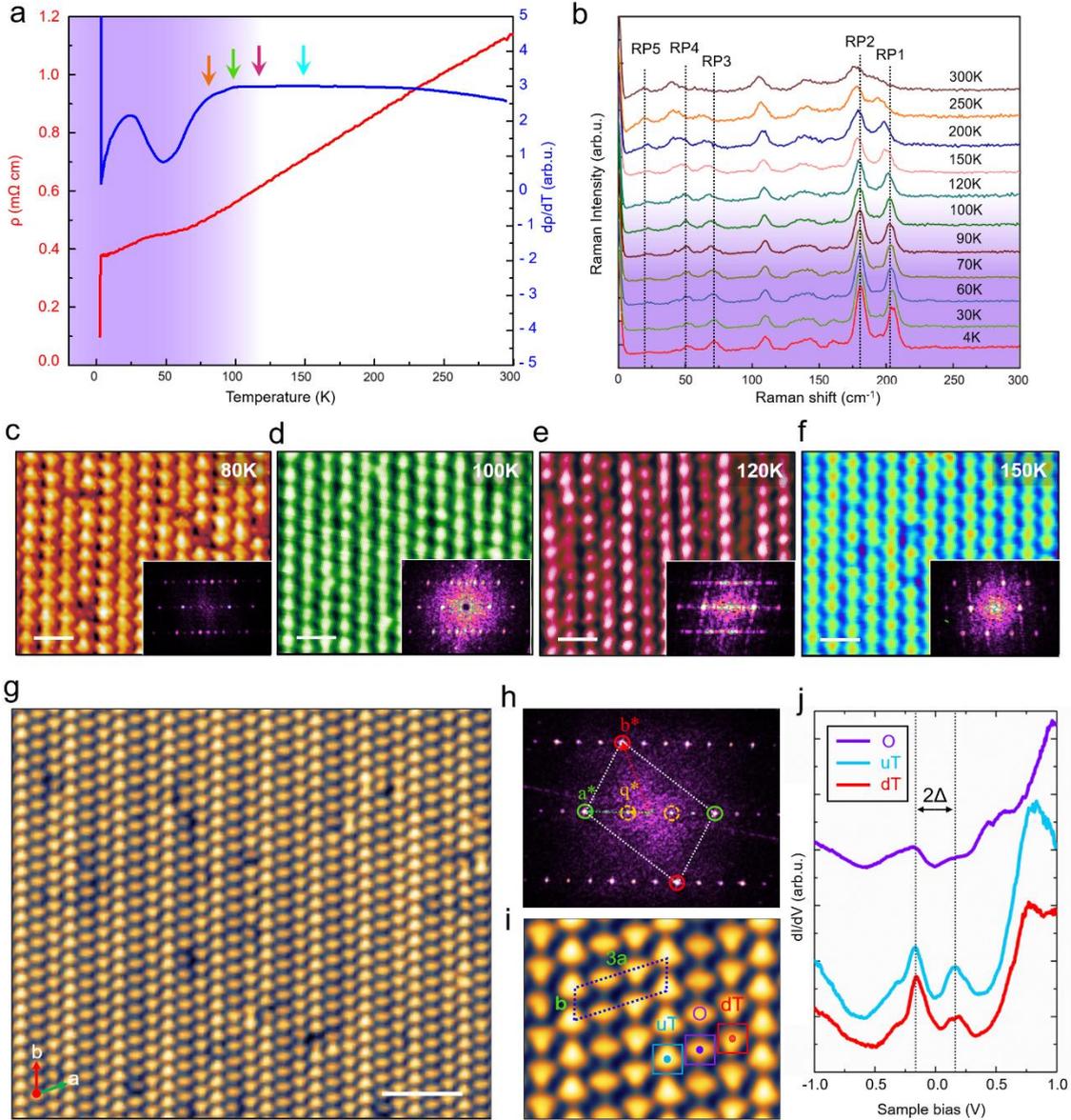

FIG. 2. Charge order transitions in the ATS super-atomic crystal. (a) Temperature-dependent resistivity (red) and its derivative (blue) curves. (b) Temperature-dependent Raman spectrum. Five temperature-dependent Raman peaks were marked as RP1-5. (c–f), STM topographies and corresponding FFT patterns obtained at 80 K (c), 100 K (d), 120 K (e) and 150 K (f), which are marked by the color-coded arrows in (a). (g) Large-scale STM topography at 9 K. (h) The corresponding FFT of (g). The green and red circles mark the Bravais vectors ($a^*$ and $b^*$) of the pristine *ab*-plane layer, respectively. The orange circles mark the Bravais vectors ($q^* = a^*/3$) of the triple stripe charge order. (i) High-resolution STM image for the 3 × 1 supercell of the triple-cube CDW. (j) STS

spectra of the uT, O and dT cube units, showing a charge order gap of ~0.17 eV. (g) Scale bar, 5 nm, Vs = +1.2 V; (h) Scale bar, 2 nm, Vs = +1.2 V.

In Fig. 2(i), an enlarged image depicts the additional order of sequentially emergent electrical polarization along the *b*-axis (Fig. S10) [43]. Particularly, for each stripe period, three chains of the ATS cubes show "up-triangular" (uT), "olive" (O) and "down-triangular" (dT) shapes in the occupied state, respectively. Figure 2(j) shows the chain-specific conductance (d$I$/d$V$) spectra obtained on the uT, O, and dT chains. All these show a quasi-gap of ~0.17 eV around the Fermi level below 120 K (Fig. S11), suggesting their CDW characteristics, referred to as triple-cube CDW (tc-CDW) hereafter. The formation of tc-CDW only in the *a*-axis direction of the ATS crystal does not open a full gap near the Fermi level in each of these spectra, where the density of states (DOSs) are incompletely suppressed. These partially suppressed gaps suggest the preserved metallicity of the ATS crystal in the presence of tc-CDW, which is confirmed by the transport measurements (see Supplementary Note 2 for more details) where an obscure feature could be identified only in the secondary differential resistance curve around 100~120 K (Fig. S8b).) Observations of this transition at 120 K using both STM imaging (a surface characterization technique), and Raman and electron transport measurements (two bulk measurement techniques), indicate that the tc-CDW transition occurs both on the surface and in the bulk of the ATS bulk crystal. No apparent temperature-dependent hysteresis has been observed in the above experimental measurements (see Fig. S12 for more details).

### C. Electronic structures of ATS and likely origin of tc-CDW

Two types of charge orders sequentially emerged around 90~120 K, namely the tc-CDW order along the *a*-axis and an electrical polarization along the *b*-axis (Fig. S13). Figure 3(a) illustrates the 2D Brillouin zone (BZ) of a cleaved ATS surface, superimposed with its real-space lattice vectors and an atomic model. Figure 3(b) depicts its electronic band structures along high-symmetric directions. Bands 1–4 appear nearly flat along $\Gamma$ - $X$, whereas they are highly dispersive and cross the Fermi level ($E_F$) along $\Gamma$ - $Y$. This explicitly indicates the anisotropy of the inter-cube interactions. These four bands correspond to the inter-cube bonding states along the *b*-axis, which have been hybridized between four pairs of Te atoms on the two facing planes of two adjacent cubes (Fig. S6). Band 5 (orange) intersects bands 3 (light green) and 4 (light pink), which opens two small gaps of ~10 meV around points 1/3 ($\Gamma$ - $X$) and 2/3 ($\Gamma$ - $X$) in the 2D BZ, respectively, and leaves a flat band nearly sitting at the $E_F$ between them. Its bandwidth along $\Gamma$ - $X$ approaches 0.97 eV but is only 0.50 eV in the $\Gamma$ - $Y$ direction, indicating stronger across-chain inter-cube interactions along $\Gamma$ - $X$. The anisotropy of inter-cube interactions in band 5 was confirmed by visualizing its wavefunction norm [Fig. S6(i)], depicting an inter-chain Te…Te bonding state and nearly isolated ATS cubes within the chain.

These five bands are, thus, categorized into two groups of highly anisotropic bands according to the plotted band structures and visualized wavefunction norms, which are illustrated in the plot of Fermi surface (FS) shown in Fig. 3(c). The Fermi surface is plotted with a smearing energy of 16 meV. Fermi surface sheets, comprising bands 1–4, are parallel to $\Gamma$ - $X$ and intersect at different points with a large vertically aligned ellipse originating from band 5. At those intersections, multiple "hotspots" [44] were formed, exhibiting bandgaps and embedding a small closed ellipse comprising bands 3/4/5. The hotspots are connected by vectors that are nearly parallel to the $\Gamma$ - $X$ direction (red arrows), and their lengths are very close to 1/3 ($\Gamma$ - $X$).) We also calculated the electronic band structure and FS of a monolayer model and present these results in Fig. S7. It exhibits comparable electronic structures, that is, bands 1–5 were highly anisotropic, and intersect Fermi surface sheets and ellipse, which indicates weak interlayer electronic coupling in ATS crystals.

The tc-CDW may be a result of electron–phonon coupling and/or electron correlation. Our DFT calculations reveal no appreciable atomic displacements accompanied by the tc-CDW along the *a*-axis in both the bulk and monolayer models, indicating a predominant electronic nature of the tc-CDW. This electronic nature is also further confirmed by the absence of Kohn anomaly [45] (Fig. S14) revealed using DFT, nearly identical chain distances observed in our STM imaging (Fig. S15), and inappreciable shadow bands induced by band-folding in our ARPES spectra (Fig. 3). This implies that electron correlation likely plays a major role in forming the tc-CDW. Strong electron–electron interactions between band 5 and bands 1–4, and/or within band 5, were evidenced by the vectors connecting hotspot pairs, and the large joint density of states (JDOS) originated from the intersecting region. Most corresponding vectors are very close to 1/3 ($\Gamma$ - $X$). Thus, the observed tc-CDW is, most likely, due to strong inter-chain Coulomb interaction across ATS chains through Te…Te quasi-bonding in the middle sublayer and significant electron hopping along ATS chains (the $b$-axis) through Te…Te noncovalent interactions in all sublayers. In addition, those two categories of anisotropic bands (i.e., bands 1–4 versus band 5) indicate stronger inter-cube interactions within the chains (along the $b$-axis). Therefore, each chain on the surface can be qualitatively regarded as one supersite and the surface could thus be modeled in a 1D FS nesting picture [46-48] along the $a$-axis (Fig. S16), explaining the tendency to form a quasi-1D CDW along that direction.

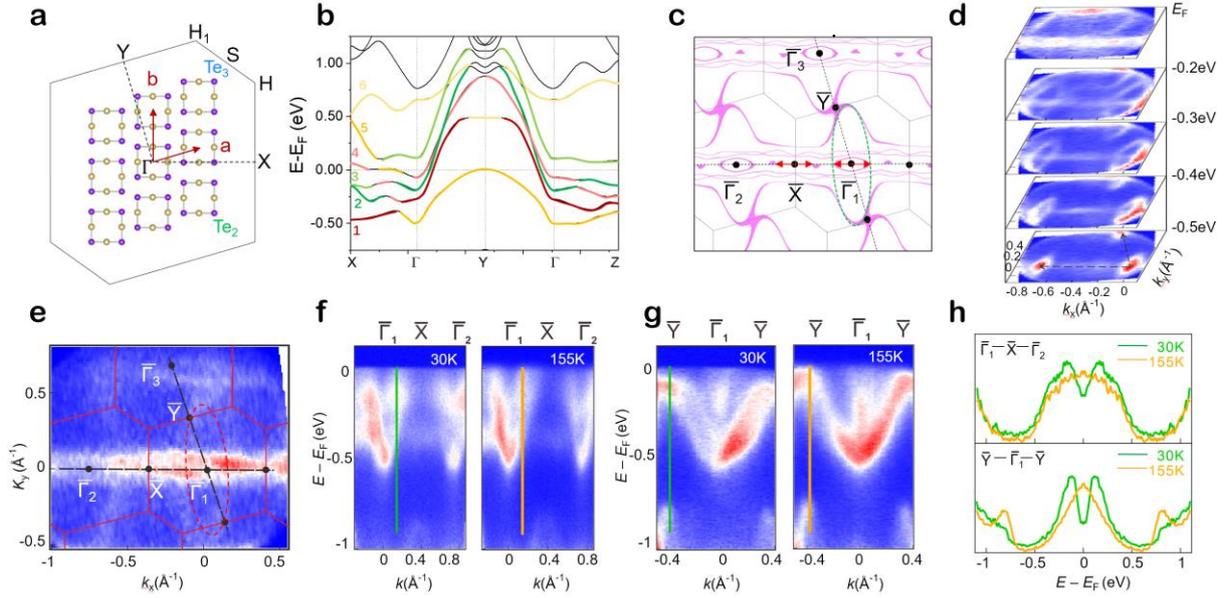

FIG. 3. Electronic structure of the ATS crystal. (a) Two-dimensional Brillouin zone (BZ) of an ATS crystal surface (the $ab$-plane) and the corresponding structural model in real space. $a$ and $b$ represent the lattice vectors. (b) Theoretical band structure of bulk ATS along high-symmetric lines. Bands 1–6 were numbered according to their orders of eigen-energies at the $X$ point (Fig. S6). (c) Constant energy contours in the 2D Brillouin zone ($k_z = 0$) at the adjacent Fermi level with an energy broadening of ±8 meV. The red arrows indicate two typical nesting vectors relevant with the tc-CDW. (d) Constant-energy intensity plots with respect to the Fermi level obtained from ARPES measurements. Two high symmetry paths ($\Gamma$ - $X$ - $\Gamma$ and $\Gamma$ - $Y$ - $\Gamma$) are indicated using two black dashed lines at -0.5 eV plane. (e) Integrated intensity plot with $E_F \pm 40$ meV. The large elliptical FS was marked using a red dashed line. (f, g) Two-dimensional intensity plots measured along $\Gamma_1$ – $X$ - $\Gamma_2$ (f) and $Y$ - $\Gamma_1$ - $Y$(g) at T = 30 K (left) and T = 155 K (right), respectively. (h) The symmetrized energy distribution curves at 1/3 ($\Gamma$ - $X$) and the $Y$ point from (f) and (g), as measured at 30 K (green) and 155 K (orange), respectively.

While our DFT results indicate that both strong Coulomb interactions and significant electron hopping could be utilized through noncovalent interactions in the ATS layer, the predicted electronic structures were experimentally verified using ARPES measurements. Figure 3(d) shows constant-energy plots starting at $E_F$ and decreasing by 0.5 eV below $E_F$ at T = 30 K. The figure shows two converged points in the intensity map acquired at -0.5 eV, which correspond to the $\Gamma_1$ and $\Gamma_2$ points of two adjacent BZs, respectively. Determination of the $\Gamma$ point position allows for the further derivation of the positions of high symmetry lines $\Gamma$ - $Y$ and $\Gamma$ - $X$, and the BZ boundary. Figure 3(e) shows the intensity map at $E_F$, where the 2D BZ boundary is highlighted in red solid lines. The large elliptical Fermi surface (indicated by the red dashed lines) and the relatively flat Fermi sheets along $\Gamma$ - $X$ were clearly resolved, which is highly consistent with the calculated FS shown in Fig. 3(c).

Band dispersion measurements were performed at $T$ = 30 and 155 K, to clarify the origin of the partial gap found in STS. Two cuts along high symmetry lines $\Gamma_1$ - $X$ - $\Gamma_2$ and $Y$ - $\Gamma_1$ - $Y$ are presented in Fig. 3(f) and 3(g), respectively. Along $\Gamma_1$ - $X$ - $\Gamma_2$, a dispersive band (band 5) crosses the Fermi level ($E_F$) near 1/3 ($\Gamma$ - $X$) at 155 K and an electronic bandgap is opened around $E_F$ at 30 K. Along $Y$ - $\Gamma_1$ - $Y$, band 5 reaches the Fermi level near the $Y$ point at 155 K, but it opens a bandgap around $E_F$ at 30 K. The gaps along the two lines were more clearly revealed using symmetrized energy distribution curve cuts, as shown in Fig. 3(h). The determined gap size is ~135 meV (at 30 K), comparable with the STS result of ~0.17 eV (at 9 K).

### D. Interweaving polar electronic states

Electrical polarization along the *b*-axis was clearly resolved in STM imaging, which breaks the degeneracy of inter-cubic Te…Te interactions within ATS chains that reduces the JDOS near the $Y$ point (Fig. 3(b) and 3(c)). Figure 4(a) and 4(b) show high-resolution STM images of the filled and empty states of the (uT-O-dT) tricube, respectively. Remarkably, the directions of each triangular chain (uT & dT) are oriented opposite to each other in the filled- and empty-state images. This indicates that the polarization along the *b*-axis is relevant with electronic hybridization, whereas no polarization occurs in the cubes of O chains. Figure 4(c–e) provides a cartoon showing the formed electrical dipoles, which are ascribed to charge polarization within the uT and dT cubes along the *b*-axis. The dipoles are parallel aligned within the uT and dT chains but antiparallelly oriented between uT and dT chains (Fig. S17).

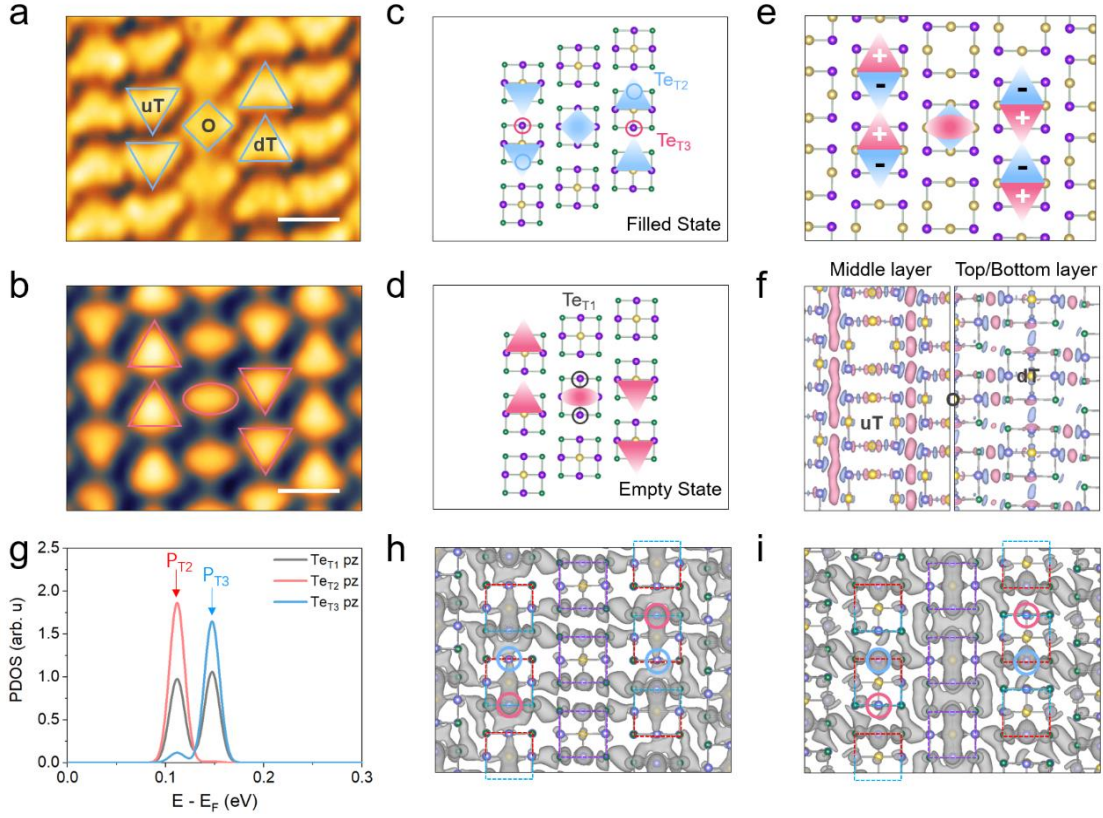

FIG. 4. Anti-polar electronic states of ATS cubes. (a, b) STM topographic images of the ATS cubes at filled (a) and empty states (b), respectively. (a) scale bar 1nm, Vs = +1.2V; (b) scale bar 1nm, Vs = -1.2V. They show inter-chain bonding- (c) and antibonding-like (d) states among the (uT-O-dT) tri-cube. (e) Illustration of polarized electronic states at filled and empty states. (f), Inter-chain differential charge density (DCD) of an ATS monolayer where red and blue contours represent charge accumulation and depletion, respectively. Slabs of middle and top/bottom are shown in the left and right panels, respectively. The isosurface value is 0.0005 e/Bohr$^3$. (g), Projected density of states (PDOS) of $p_z$ orbitals of three types of Te atoms, as denoted in (c) and (d), in the top sub-layer. (h,i), Wavefunction norms of states $PT_2$ (h) and $PT_3$ (i), respectively, where the dashed boxes outline ATS cubes.

Appreciable accumulation of (depleted) DOS was found at the uT-O and O-dT inter-chain regions in the filled-state (empty-state) image {Fig. 4(a) [4(b)]}, pointing toward the essential roles that the inter-chain electronic hybridization plays in inducing these polarized electronic states. Figure 4(f) depicts inter-chain differential charge density (DCD) in the middle (left panel) and top/bottom (right panel) sublayers of an ATS monolayer in the triple-cube stripe configuration. The middle sublayer engenders strong inter-chain interactions where covalent-like Te…Te hybridizations, that is, charge reduction (blue) near Te atoms and charge accumulation (red) in between them, were identified in the inter-chain region. On the top/bottom sublayer, we observe enhanced inter-chain Te…Se covalent-like characteristics and slightly weakened intrachain Te…Te interactions, as reflected by charge accumulation in the inter-chain areas and charge reduction at the inter-cube region within chains, respectively.

Electronic interactions, primarily hybridization, are capable of introducing spatially polarized distribution of electronic states along the *b*-axis in the uT and dT chains, which are clearly evidenced by the projected DOS (PDOS) of Te atoms at the top sublayer (Fig. 4(g–i) and Fig. S13). Compared to the Te$_{T1}$ atom from the non-polarized O-chain, atoms Te$_{T2}$ and Te$_{T3}$ of the uT and dT chains display apparent inversion-symmetry breaking. Representative peaks P$_{T2}$ and P$_{T3}$, separated by 35 meV, are both distributed on Te$_{T1}$ and preferably located on Te$_{T2}$ and Te$_{T3}$, respectively, showing significant electrical polarization in the uT and dT chains. Real-space wavefunction norms of these two peaks are visualized in Fig. 4(h) and 4(i), which display strongly polarized states (inversion-symmetry breaking) along the *b*-axis in the uT and dT chains and inversion-symmetric states for the O-chain.

## III. DISCUSSION

As discussed in detail elsewhere [49], the tc-CDW (along the *a*-axis) coexists with the emergent superconductivity (below ~3K) under ambient pressure, but they compete under high pressures. This also essentially enhances the uT-O and O-dT across-chain interactions, which strengthens Te…Te electronic hybridizations in the across-chain direction but leaves the inversion symmetry maintained for the O-chain cubes. This reinforced hybridization tends to reduce the high JDOS near the Y point and to lift the degeneracy of inter-cube electronic states within chains. Consequently, polarized electronic states (along the *b*-axis) are formed at low temperature. An antiparallel polarized configuration is thus locked to maintain the inversion symmetry of the O cube. In other words, specific anti-polarization configurations are selectively determined by the tc-CDW charge order, which explains the use of the term "interweaving polar charge" orders.

Spatially polarized metallic states were originally proposed by P. W. Anderson and E. I. Blount in 1965 [50]. Until now, roughly 40 types of polar metallic materials have been theoretically predicted [51]; however, only a few of these have been discovered and experimentally confirmed [52-54]. Nevertheless, an anti-polar metallic state was discovered in this study, in which the anti-polar state was primarily uncovered using STM imaging in the real-space while the metallic state was verified in both cube-resolved STS and macroscopic electrical transport measurements. The anti-polar metallic state was primarily ascribed to strong electronic interactions rather than spatial displacements of atoms.

In summary, we demonstrated that strong electron–electron interactions are at play in a non-covalently bonded layered super-atomic crystal. Such exceptionally strong interaction leads to the tc-CDW, the spatially polarized electronic states, and their interweaving anti-polar configuration coexisting with metallic states. These findings reveal that super-atomic crystals, comprising atomic clusters through noncovalent interactions, are suitable for the exploration, manipulation, and utilization of exotic electronic properties, which were traditionally investigated in covalently bonded solids. Given the high tunability of super atoms, more super-atomic solids and layers can be proposed, prepared, or even rationally designed by tailoring their super-atomic building blocks and the inter-block interactions among them. This allows for the exploration of those parameter regimes and thus the observation of emergent phenomena that were inaccessible in conventional quantum materials.

## IV. MATERIALS AND METHODS

### A. Sample preparation and STM measurements

Single crystals of $Au_6Te_{12}Se_8$ (ATS) were grown by the self-flux method. They are in the shape of small platelets with shining mirror-like surfaces, typically $2 \times 4 \times 1$ mm$^3$. Samples were cleaved in ultrahigh vacuum at room temperature and subsequently cooled for STM measurements, which were performed in a commercial variable-temperature STM (PanScan Freedom, RHK) operated in ultrahigh vacuum. Electrochemically etched polycrystalline tungsten calibrated on clean Au(111) surfaces was used for all our STM measurements tips. The STM topography was acquired in the constant-current mode, and the d$I$/d$V$ spectra were collected using a standard lock-in technique with a modulation frequency of 999.1 Hz. STM measurements were performed mostly at 9 K for high-resolution imaging, and at variable temperatures (from 9 K to room temperature) for the phase transition characterization.

### B. XRD, SEM, and transport measurements

Powder X-ray diffraction (PXRD) patterns of polycrystalline ATS were measured using a panalytical X'pert diffractometer with the Cu-K$_\alpha$ anode ($\lambda$ = 1.5408 Å). Scanning electron microscopy (SEM) images of the ATS single crystal were acquired using a Hitachi S-4800 FE-SEM. Electrical resistivity ($\rho$) was measured using the standard four-wire method in a PPMS (Quantum Design).

### C. Raman measurements

Raman spectroscopy measurements were performed using a home-built low-wavenumber and variable-temperature Raman system equipped with a semiconductor laser ($\lambda$=532 nm), 50X objective (numerical aperture 0.8), and a 600 lines mm$^{-1}$ grating. The sample was placed in a cryostat (attocube800 systems AG, Germany). To prevent potential CDW phase transitions induced by laser irradiation, the laser irradiance was kept below 100 μW/μm$^2$. Low-wavenumber Raman filters (Ondax Inc., USA) were used to achieve a cut-off Raman shift down to ~10 cm$^{-1}$. The step size of Raman mapping was 1 μm. All the peaks were calibrated with the Si peak at 520.7 cm$^{-1}$.

### D. ARPES measurements

ARPES measurements were performed at Renmin University of China using a Scienta DA30 analyzer and at a photon energy of 10.05 eV, as well as at the BL13U beamline of National Synchrotron Radiation Laboratory (NSRL) equipped with a Scienta R4000 analyzer. The energy and angular resolution were set to 10 meV and 0.3°, respectively. Clean surfaces for ARPES measurements were obtained by *in situ* sample cleaving. Photoemission spectra presented in this study were recorded at $T$ = 30 K and 155 K using the photon energy from 21 to 45 eV in a working ultrahigh vacuum better than 6 × 10$^{-11}$ Torr.

### E. Theoretical calculations

Density functional theory calculations were performed using the generalized gradient approximation for the exchange-correlation potential, the projector augmented wave method [55, 56], and a plane-wave basis set as implemented in the Vienna *ab-initio* simulation package (VASP) [57] and Quantum Espresso package (QE) [58]. Dispersion corrections were made at the van der Waals density functional (vdW-DF) level [59, 60], with the optB86b functional for the exchange potential (optB86b-vdW) [61, 62]. The kinetic energy cut-off for the plane-wave basis was set to 700 and 350 eV for all geometric property and electronic structure calculations, respectively. Two $k$-meshes of $7 \times 7 \times 5$ and $7 \times 7 \times 1$ were used to sample the first Brillouin zone in structure optimizations of the bulk and monolayer $Au_6Te_{12}Se_8$ (ATS) crystals, respectively. The $k$-meshes were increased to $15 \times 15 \times 15$ and $15 \times 15 \times 1$ in electronic structure calculations. An even denser $k$-mesh of $30 \times 30 \times 1$ was used to plot the 2D Fermi surface of the bulk ($k_z = 0$) and monolayer crystals. Density functional perturbation theory [63] was employed to calculate the vibrational frequencies of the ATS bulk crystal at the $\Gamma$ point and point $\boldsymbol{q}_{\text{tc-CDW}} = (1/3,0,0)$ in the reciprocal space using the VASP and Quantum Espresso packages, respectively. A $3 \times 1$ supercell was used to model the ATS monolayer in the tc-CDW state. The Brillouin zone was sampled using a $2 \times 7 \times 1$ $k$-mesh for both structural optimizations and electronic structure calculations. A vacuum layer of 17 Å was used to eliminate image interactions among adjacent supercells. The shape and volume of the supercell and all atomic positions were fully relaxed until the residual force per atom was less than 0.01 eV/Å. A Methfessel–Paxton smearing of 0.05 eV and the Bloch corrected tetrahedron method were used for Brillouin zone integral in calculations of geometric and electronic structures, respectively. A Gaussian smearing of 0.01 eV was used to plot the projected density of states for the $3 \times 1$ supercell. We calculated charge densities of three single ATS chains (i.e., $\rho_{uT}$, $\rho_O$ and $\rho_{dT}$) using the same geometry and precision as those in the $3 \times 1$ supercell and subtracted them from total charge density of the supercell ($\rho_{total}$), thereby obtaining the inter-chain differential charge density ($\rho_{DCD}$), i.e., $\rho_{DCD} = \rho_{total} - \rho_{uT} - \rho_O - \rho_{dT}$.

# ACKNOWLEDGMENTS

This project is supported by the Ministry of Science and Technology (MOST) of China (No. 2018YFE0202700, 2016YFA0200700), the National Natural Science Foundation of China (NSFC) (No. 61674045, 11604063, 61911540074, 51922105, 11774421, 61761166009, 11974422), the Strategic Priority Research Program (Chinese Academy of Sciences, CAS) (No. XDB30000000), and the Fundamental Research Funds for the Central Universities and the Research Funds of Renmin University of China under grant numbers 21XNLG27 (Z.C.) and 22XNKJ30 (W.J.). L.W. was supported by the Outstanding Innovative Talents Cultivation Funded Programs 2022 of Renmin University of China. Calculations were performed at the Physics Lab of High-Performance Computing of Renmin University of China, Shanghai Supercomputer Center.

Z.C., S.W., J.G. and W.J. conceived the research project. S.X. L.L., Q.Z., X.L., L.H., H.J.G. and Z.C. performed the STM experiments and analysis of STM data. Z.W. and S.W. performed the ARPES experiments and analysis of ARPES data. X.C., J.G. and X.C. grew the single crystals and performed transport, SEM and XRD measurements. H.L., S.H., S.C. and L.X. performed Raman measurements. L.W., L. Z. and W.J. performed the DFT calculations. S.X., L.W., Z.W, Z.C. S.W. and W.J. wrote the manuscript with inputs from all authors.

# Supplemental Information

# Interweaving Polar Charge Orders in a Layered Metallic Super-atomic Crystal

Shuya Xing[1,+], Linlu Wu[1,+], Zilu Wang[1,+], Xu Chen[2,+], Haining Liu[3,4], Shuo Han[1], Le Lei[1], Linwei Zhou[1], Qi Zheng[2,4], Li Huang[2,4], Xiao Lin[4], Shanshan Chen[1], Liming Xie[3,4], Xiaolong Chen[2,4,5] Hong-Jun Gao[2,4], Zhihai Cheng[1,*], Jiangang Guo[2,5,*], Shancai Wang[1,*], and Wei Ji[1,*]

[1]*Beijing Key Laboratory of Optoelectronic Functional Materials & Micro-nano Devices, Department of Physics, Renmin University of China, Beijing 100872, China*

[2]*Beijing National Laboratory for Condensed Matter Physics, Institute of Physics, Chinese Academy of Sciences, P.O. Box 603, Beijing 100190, China.*

[3]*CAS Key Laboratory of Standardization and Measurement for Nanotechnology, CAS Centre for Excellence in Nanoscience, National Centre for Nanoscience and Technology, Beijing 100190, China*

[4]*University of Chinese Academy of Sciences, Beijing 100039, China*

[5]*Songshan Lake Materials Laboratory, Dongguan, Guangdong 523808, China*

[+] These authors contributed equally: Shuya Xing, Linlu Wu, Zilu Wang, Xu Chen.
* Email: zhihaicheng@ruc.edu.cn    jgguo@iphy.ac.cn    scw@ruc.edu.cn    wji@ruc.edu.cn

# Supplementary Note 1: "design process" of super-atomic crystal

Emergence that relates to symmetry breaking is one of the core concepts in condensed matter physics. In this work, we intended to explore potential emergent electronic structures and electrical properties in materials assembled from atomic clusters through non-covalent bonding. To achieve this goal, we have to find a cluster that satisfies the following requirements:

(i) **Thermal-stability**: many clusters are stabilized by charging or/and ligands. Charged clusters are usually assembled with ions of opposite charges into ionic crystals. These crystals largely show bandgaps around the Fermi level and have difficulties to render highly degenerate electronic states near the Fermi level for inducing some emergent electronic structures by strong electron-electron interactions. For a similar reason, we have to avoid using ligands as they greatly suppress inter-cluster electronic interactions. Therefore, we have to consider those clusters that are neutral and self-saturated, e.g. buckyballs, TM@$X_{12}$ (TM=transition metal, X=Si/Sn), and ATS.

(ii) **High symmetry**: Emergent electronic structures are typically relevant with symmetry breaking. We should thus consider a candidate of high geometric symmetry. Buckyballs, like $C_{60}$, constitute a type of spherical clusters showing very high geometric symmetry. But this is a double-edged sword that makes it more challenging to obtain well defined inter-cluster interactions. A cubic cluster, like ATS, shows reasonably high symmetry and could offer robust face-to-face inter-cluster interaction, which well balances the requirements of high symmetry and well-defined inter-cluster interactions.

(iii) **High density of states near the Fermi level**: A large HOMO-LUMO gap could substantially stabilize a cluster, but would be contradictory with the required high electron density near the Fermi level of its assembly for enabling strong electron-electron interactions. Therefore, we may need a "combined" cluster in which some of the atoms provide a firm atomic frame whereas some others offer the electronic states. ATS, by combining three elements into a cluster, fit this requirement. While the Se-Te, also Te-Au, sigma bonds offer a firm atomic frame, six MOs, primarily composed of extended out-of-plane Te *p* orbitals sit around a small (0.22 eV) HOMO-LUMO gap.

(iv) **Appreciable inter-cluster non-covalent interaction**: Inter-cluster covalent bonding often results in strong electronic hybridization of those states around the HOMO-LUMO gap of individual clusters. Such hybridization, accompanied by significant energy level splitting, destroys the designs of the individual cluster. In our recent studies, we found a type of interaction termed covalent-like quasi-bonding (CLQB), existing in inter-layer and inter-cluster interactions of many 2D layers and $C_{60}$ clusters, respectively. As we demonstrated in our previous studies, Te…Te [1] and Se…Se [2] CLQBs lead to significant but gentle inter-chain electronic interactions. Therefore, we considered Te and Se elements in our design of the target cluster.

While reviewing the literature, we found a JCP paper [3] reporting a simple cubic structure formed by $Au_{1-x}Te_x$ (Au 15-40%, $0.6 < x < 0.85$), in which Au and Te randomly occupy the eight corners of the simple cubic cell. Doping higher-electronegativity Se atoms into the AuTe cubes, results in an ordered arrangement of Au, Te, and Se atoms in a cubic cluster, namely the ATS cube [4]. An assembly of the ATS clusters shows significant inter-cluster (face-to-face) Te…Te, Te…Se and Se…Se CLQBs whose hybridized states reside around the Fermi level, which means that our design goal is thus achieved.

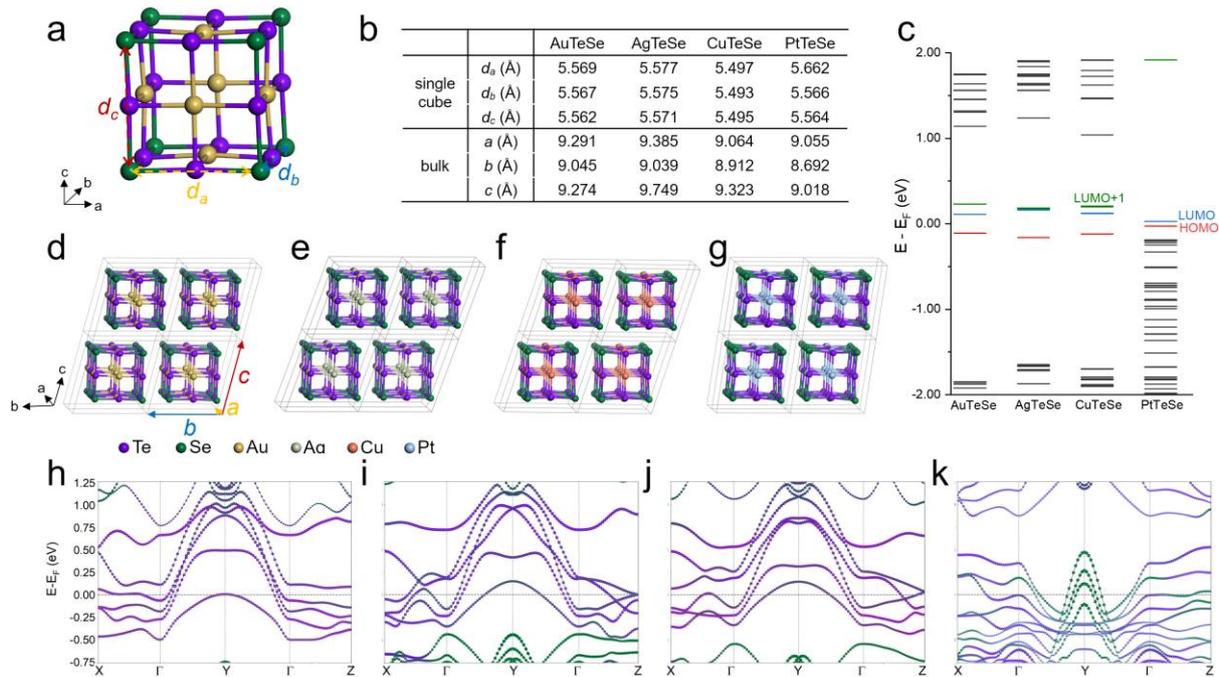

**FIG S1:** Geometric and electronic structures of AuTe$_2$Se$_{4/3}$, AgTe$_2$Se$_{4/3}$, CuTe$_2$Se$_{4/3}$, and PtTe$_2$Se$_{4/3}$ single cubes and bulk crystals. (a) Structural model of the cube. The Se-Se distances along three directions are labeled by orange ($d_a$), blue ($d_b$) and red ($d_c$) dash arrows, respectively. (b) Size of single cube and bulk lattice constants of these four clusters. (c) Energy levels of these four clusters. $E_F$ is defined at the middle of the HOMO-LUMO gap. (d-g) (h-k) Bulk structures showing stacking configurations (electronic bandstructures) of AuTe$_2$Se$_{4/3}$ (d,h), AgTe$_2$Se$_{4/3}$ (e,i), CuTe$_2$Se$_{4/3}$ (f,j) and PtTe$_2$Se$_{4/3}$ (g,k).

We explain the "precise tunability" in detail as follows. The tuning aims to change the electronic structure of the ATS assembly in a precisely controllable way. As we mentioned in Fig. S5, Te-Se sigma bonds dominate the formation of the ATS atomic frame, which means substitution of the metal atoms does not substantially change the geometries of individual ATS cubes and thus the ATS assemblies. This feature is very attractive and is different from atomic crystals where the substitution of atoms often leads to a change in the crystal symmetry.

Given the robust geometry, we found substitution of the metal atoms finely changes the electronic structure of the cluster assemblies. As listed in (b), AuTe$_2$Se$_{4/3}$, AgTe$_2$Se$_{4/3}$ and CuTe$_2$Se$_{4/3}$ crystals share very closed cube sizes (from 5.49 to 5.58 Å) and stacking geometry, but distinctly different electronic structures. The Au based cube shows a flat band, but substitution of Au with Ag and Cu leads to a metal and slightly gapped semi-metal in the Γ-X direction of their bulk crystals, respectively, as their frontier MOs have slightly different energies in the individual cubes. This subtle and controllable change in electronic structure in the maintained geometric structure represents the "precise" tunability.

If we replace Au with Pt atoms, the PtTe$_2$Se$_{4/3}$ cluster offers very closed cube size (5.56 – 5.66 Å) and a stacking configuration in the bulk crystal highly comparable to the ATS cluster. Given the mostly maintained geometry, a substantially different electronic structure is found, showing the "high" tunability of super-atomic crystals. Besides, the dominant non-covalent interactions among clusters make the cluster assemblies rather soft. The "high" tunability is also reflected in enabling a change in inter-cluster distances or/and relative stacking configurations in a more feasible way, e.g. by using smaller energy cost or stress, and in a wider range. The manuscript arXiv:2110.10094 [5] is a pioneering work in terms of such tuning under pressure. The substitution of non-metal atoms is another strategy for mildly tuning both the geometric and electronic structures of cluster assemblies.

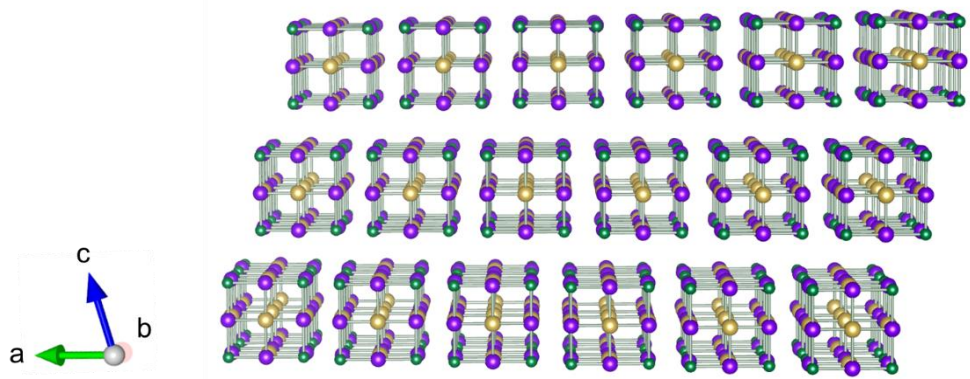

**FIG. S2:** Three-dimensional stacking structure of the layered super-atomic crystal of AuTe$_2$Se$_{4/3}$ (ATS) crystal. The 2D layers in the *ab*-plane are stacked along the *c*-axis with the relative weak Van der Waals interactions.

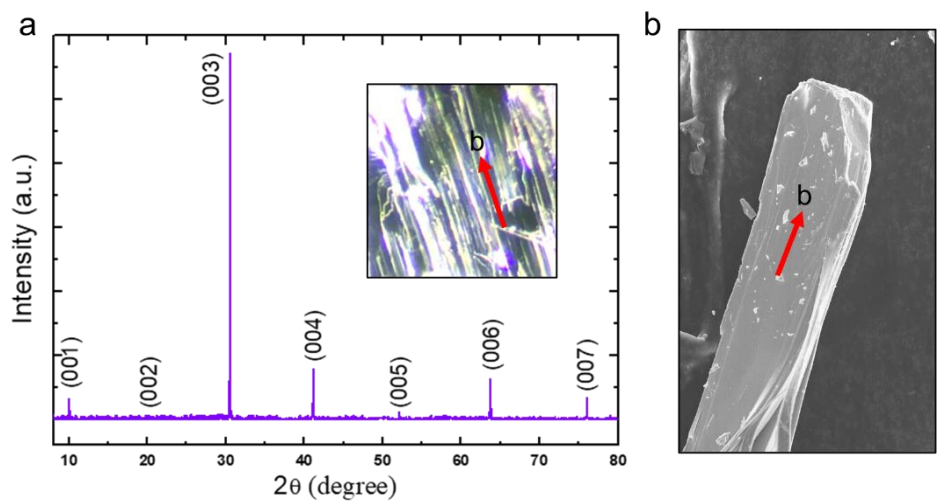

**FIG. S3**: Structural and morphological characterization of AuTe$_2$Se$_{4/3}$ (ATS) crystal. (a) The X-ray diffraction (XRD) pattern of ATS single crystal only shows [001] Bragg peaks, indicating that the *ab*-plane is the cleaving plane of ATS crystal. The inset is the optical image of the cleaved ATS single crystal. The major step direction of the cleaved plane is along the *b*-axis, reflecting the in-plane anisotropy of the *ab*-plane. (b) SEM image of a strip of ATS single crystal. The *b*-axis of the crystal is marked in both optical and SEM images.

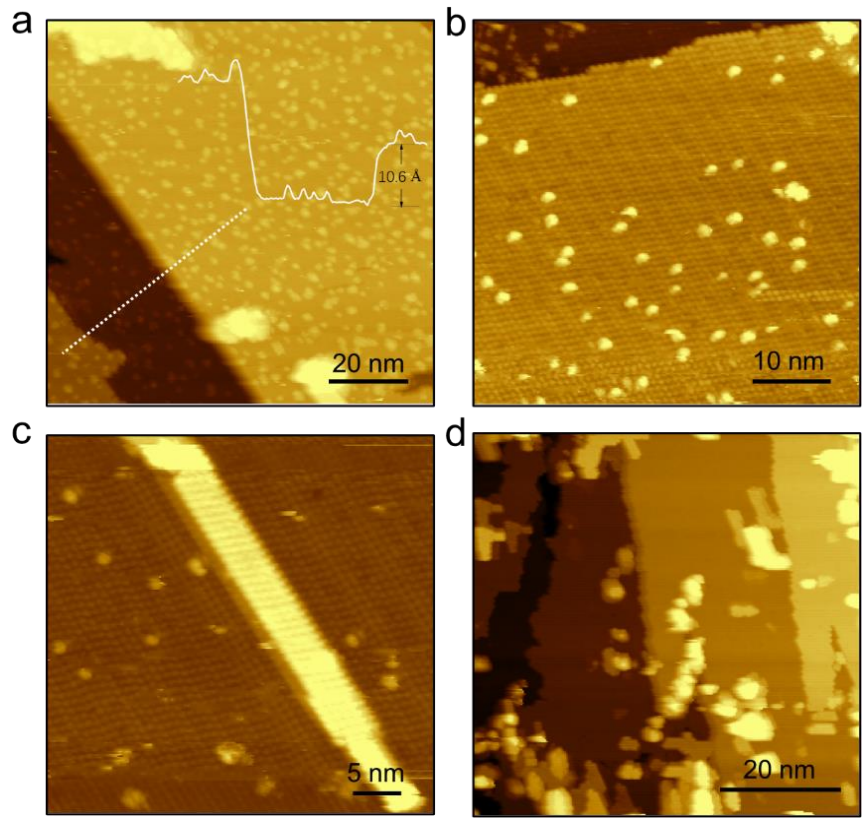

**FIG. S4:** STM topography of the cleaved ATS surface. (a) Large-scale STM image of terraces with step edges and small patches on the terraces. The line-scan profile taken along the white dotted line indicate the single- and double-layer step heights. (b) High-resolution STM image of a terrace covered by the randomly distributed zero-dimensional ATS cubes (0D). (c) High-resolution STM image of a terrace covered with a one-dimensional ATS nanoribbon (1D). (d) A typical STM image of the cleaved surface, indicating the experimental difficulty attributed to the weak inter-cube interactions. Scanning parameters are (a) $V_s$ = 2.4 V, (b) $V_s$ = 2.0 V, (c) $V_s$ = 1.6 V, (d) $V_s$ = 3.1 V. The specific 0D-1D-2D structure of the ATS crystal can be readily confirmed by the above STM results. Based on the experimental experience, there is more chance to get the clean surface by mechanical cleaving along the direction vertical to the *b*-axis.

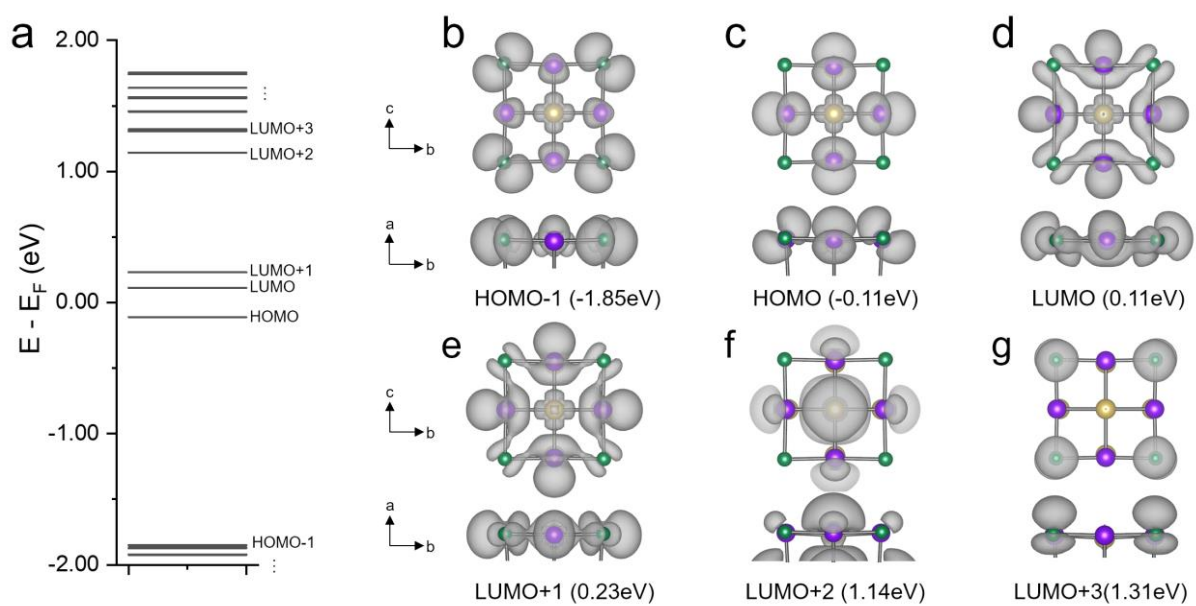

**FIG. S5:** Energy levels and visualized wavefunction norms of ATS single cube. (a) Energy levels of molecular orbitals (MOs). $E_F$ is defined as the average energy of HOMO (the highest occupied molecular orbital) and LUMO (the lowest unoccupied molecular orbital). (b-g) Wavefunction visualizations of HOMO-1 to LUMO+3 states. Among them, the HOMO, LUMO (two-fold degenerated) and LUMO+1 (three-fold degenerated) orbitals, sit around the HOMO-LUMO gap, which are relevant with those six bands near the Fermi level in the ATS monolayer or bulk. Other molecular orbitals (MOs) are over 1.0 eV away from the gap. These six MOs, forming bands 1 to 6 (see Fig. 3b, Fig. S6d and Fig. S7c), contain localized Au $d$ states. The exact shape/component of the $d$ states depends on which surface of the cube is examined (c-e). The Au $d$ states aside, HOMO is additionally comprised of out-of-plane Te $px/y/z$ orbital components, the LUMO and LUMO+1 states are similar in shape and have pronounced Te $p$ orbital components and less significant Se $p$ components. The pronounced and extended out-of-plane Te $p$ components explain the reason why Te…Te interactions dominate the inter-cluster interactions in ATS assemblies.

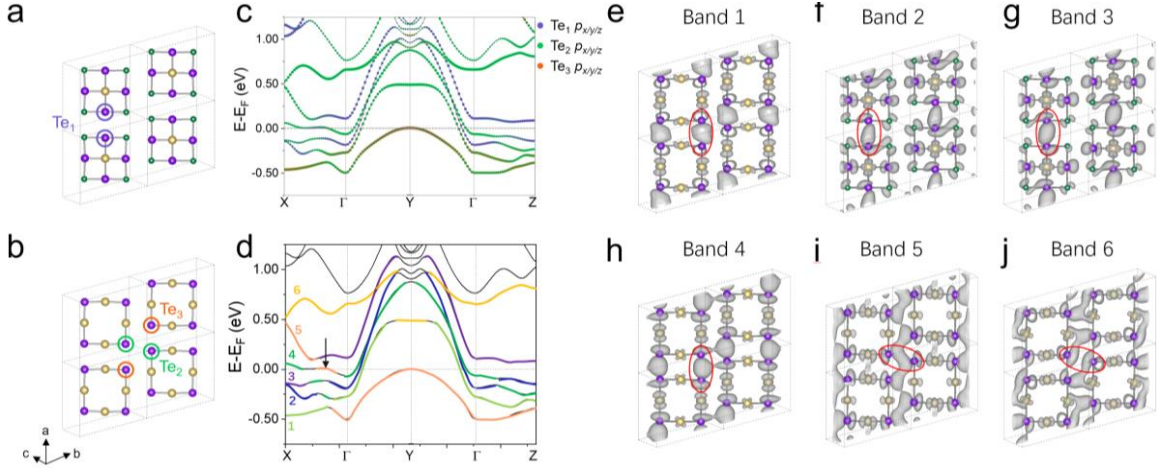

**FIG. S6:** Atomic orbital decomposed band structures and visualized wavefunction norms of bulk ATS. (a, b) The top/bottom and middle sublayers of bulk ATS. Three types of Te atoms participate in inter-cube Te…Te interactions and are labeled by purple, green, and orange circles, respectively. Unit cell of lattice is labeled by black dash line. (c, d) Electronic band structures of bulk ATS. The $p$ orbitals of Te atoms in (c) are mapped with different colors: $Te_1$-$p_{x/y/z}$, purple; $Te_2$-$p_{x/y/z}$, green; $Te_3$-$p_{x/y/z}$, orange. Bands 1–6 in (d) were numbered according to their orders of eigen-energies at the X point and color-coded by their wave-function components, as shown in (e)–(j). (e)–(j) Visualized wave-function norms of bands 1–6 at the Γ point. The wave-function components are categorized into three types, i.e., 1 and 4 (inter-cluster $Te_2$-$Te_3$ bonding states along the chain), 2 and 3 (inter-cluster $Te_1$-$Te_1$ bonding states along the chain) and 5–6 (inter-cluster $Te_2$-$Te_2$ bonding or anti-bonding states across the chain). The isosurface value is 0.001 $e$/Bohr$^3$.

The wavefunction components are categorized into three types, i.e., 1 and 5 (inter-cluster $Te_2$-$Te_3$ bonding states along the chain), 2 and 4 (inter-cluster $Te_1$-$Te_1$ bonding states along the chain) and 3 and 6 (inter-cluster $Te_2$-$Te_2$ bonding or anti-bonding states across the chain). The isosurface value is 0.001 $e$/Bohr$^3$. HOMO, LUMO and LUMO+1 are the MOs hybridizing to form bands 1-6 in the ATS monolayer and the bulk, among which Te $p_{xy}$, $p_{yz}$ and $p_{xz}$ orbitals are dominant and thus governs the inter-cluster interactions near the Fermi level (from -0.5 eV to 1.0 eV). As was shown in Fig. S5 (c)-(e), the HOMO (-0.11 eV), LUMO (0.11 eV) and LUMO+1 (0.23 eV) are primarily comprised of out-of-plane Te $p$ orbitals and localized in-plane Au $d$ orbitals.

The localized in-plane states of Au largely suppress the inter-cluster hybridization, which is otherwise promoted for the out-of-plane Te $p$ states. The hybridization of the three MOs mainly manifests as the inter-cluster Te-Te bonding and anti-bonding states [see Fig. S6 (e)-(j), and Fig. S7 (d)-(i)]. A small amount of out-of-plane $p$ orbital components of Se atoms in LUMO and LUMO+1 also contribute to the inter-cluster Te-Se interaction, but it is significantly weaker than the Te-Te interaction. Therefore, atomic orbital decomposed band structures [Fig. S6(d) and Fig. S7(b)] and $|\psi|^2$ of bands 1 to 6 [Fig. S6 (e)-(j) and Fig. S7 (d)-(i)] result in pronounced inter-cluster Te-Te bonding but nearly inappreciable Te-Au interactions.

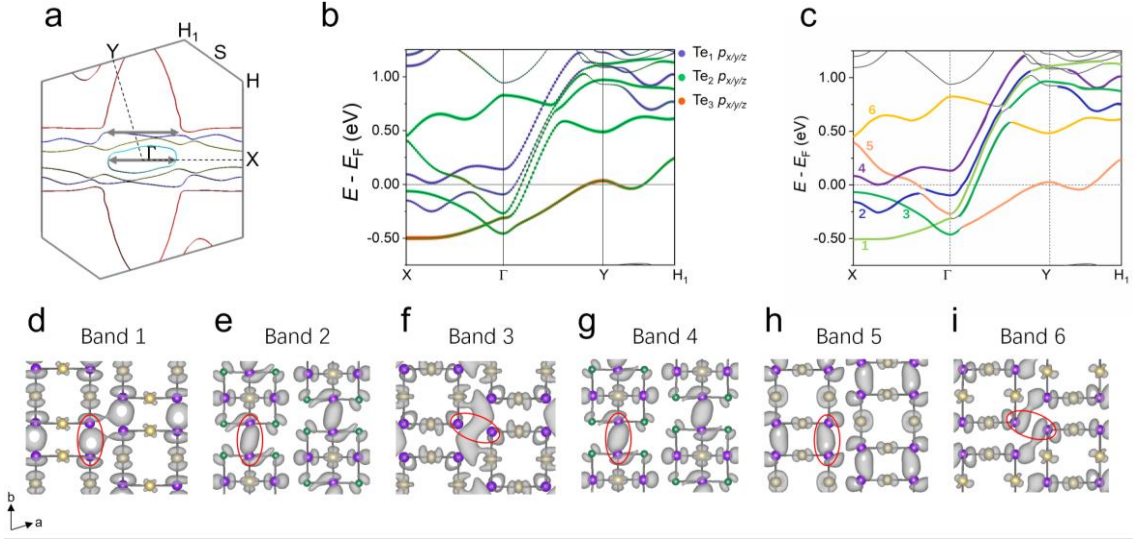

**FIG. S7:** Fermi surface, atomic orbital decomposed band dispersion and visualized wavefunction norms of pristine monolayer ATS. (a) Calculated Fermi surface of monolayer ATS with the four color-coded Fermi sheets. The possible nesting vectors of the triple-cube CDW are marked by gray arrows. (b, c) Electronic band structures of monolayer ATS. The $p$ orbitals of Te atoms in (b) are mapped with different colors: $Te_1$-$p_{x/y/z}$, purple; $Te_2$-$p_{x/y/z}$, green; $Te_3$-$p_{x/y/z}$, orange. The definition of $Te_{1-3}$ in monolayer is the same as that in the bulk, as shown in Fig. S6 (a)–(b). Bands 1–6 in (c) were numbered according to their orders of eigen-energies at the X point and color-coded according to their wavefunction components. (d)-(i) Visualized wavefunction norms of bands 1-6 at the $\Gamma$ point.

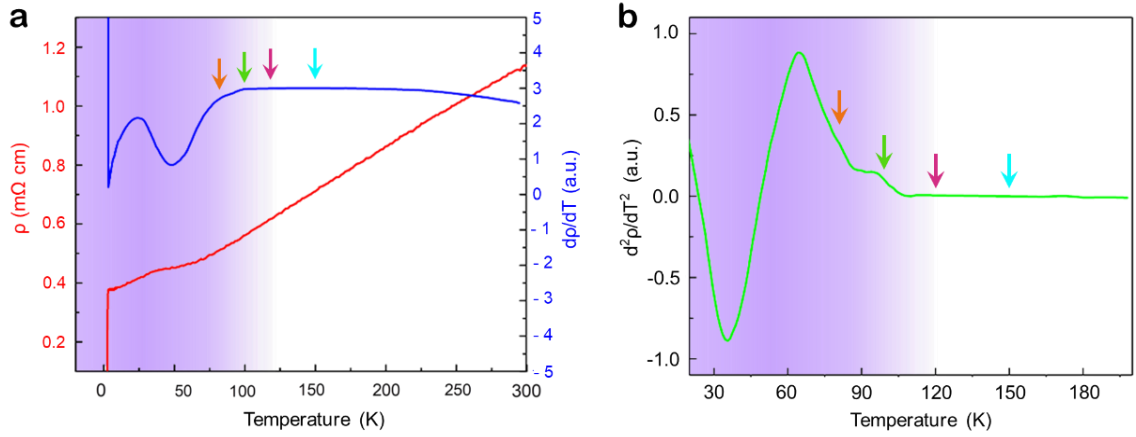

**FIG. S8:** Temperature-dependent transport measurements of the charge order transition in ATS. (a) Temperature-dependent resistivity (red) and its first-order derivative (blue) curves. (b) Second order derivative of temperature-dependent resistivity (green) curve. The kink-like feature ~90 K indicates the formation of a charge polarization, which is the anti-polar transition as revealed by STM. Another tc-CDW transition, also identified by STM is a bit too weak to show the transition, but it could be appreciable if we carefully examine the second order differential resistance at around 110 K.

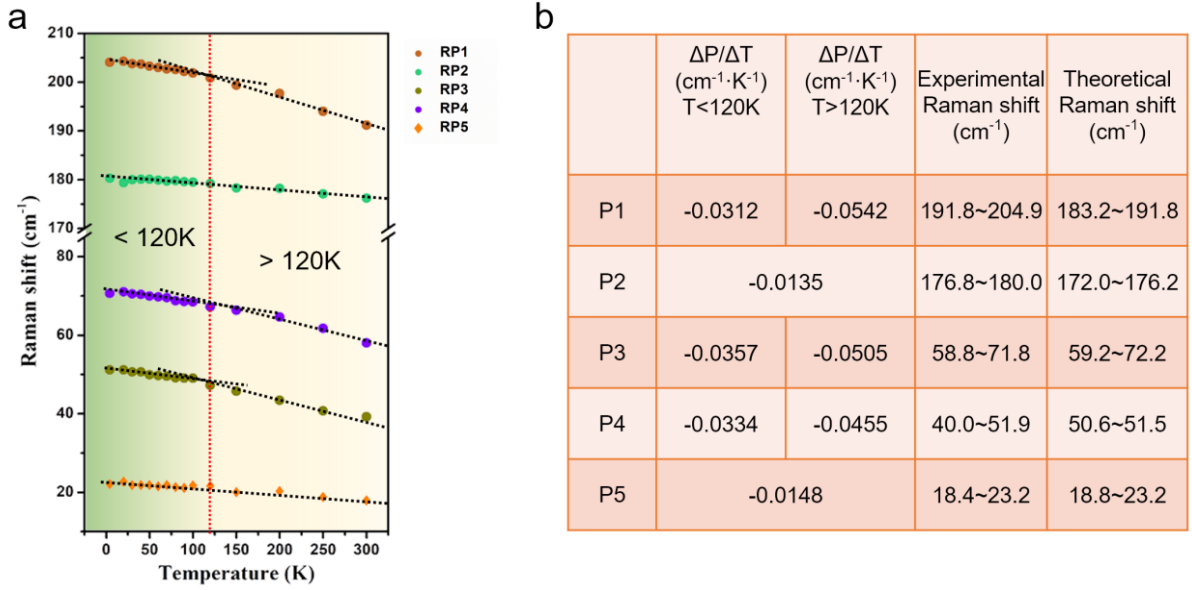

**FIG. S9:** (a) Temperature-dependent Raman shift (TDRS) frequencies for the five Raman activated modes, named RP1 to RP5 in the frequency order from high to low. The TDRS curve of each mode could be linearly fitted in one or two gradients, as listed in (b). (b) Statistical table of frequencies and linearly fitted gradients of RP1 to RP5 in (a). Here, $\Delta P$ and $\Delta T$ refer to the changes of the wave number of Raman shift peaks and temperature, respectively. For TDRS curves for modes RP1, RP3 and RP4, two gradients could be fitted in two temperature windows separately by 120 K, giving rise to the gradients of 0.031-0.036 $cm^{-1}$/K and 0.046-0.054 $cm^{-1}$/K, respectively.

Although modes RP1, RP3, and RP4 are relevant with the tc-CDW state, not all the Raman activated modes are relevant. The TDRS curves for modes RP2 and RP5 could be linearly fitted, with gradients of 0.014~0.015 $cm^{-1}$/K for the entire temperature window, i.e., from 4 to 300 K. Such shifts are usually ascribed to thermal expansion, as previously observed in many layered materials [6-8].

It was remarkable that the vibrational displacements are mainly along the *a*-axis for mode RP2, parallel to the tc-CDW direction, which evidences no structural transition accompanying the tc-CDW transition at ~120 K. Furthermore, there was no Kohn anomaly (phonon mode softening) in the vibrational frequency calculations for bulk ATS at *q* = (1/3, 0, 0) with the electronic temperature down to 15.8 K (see Fig. S14). All these results indicate that electron-phonon coupling is not the primary driving mechanism for the tc-CDW. Therefore, the tc-CDW transition occurred at 120 K is, most likely, not accompanied by any appreciable structural transition.

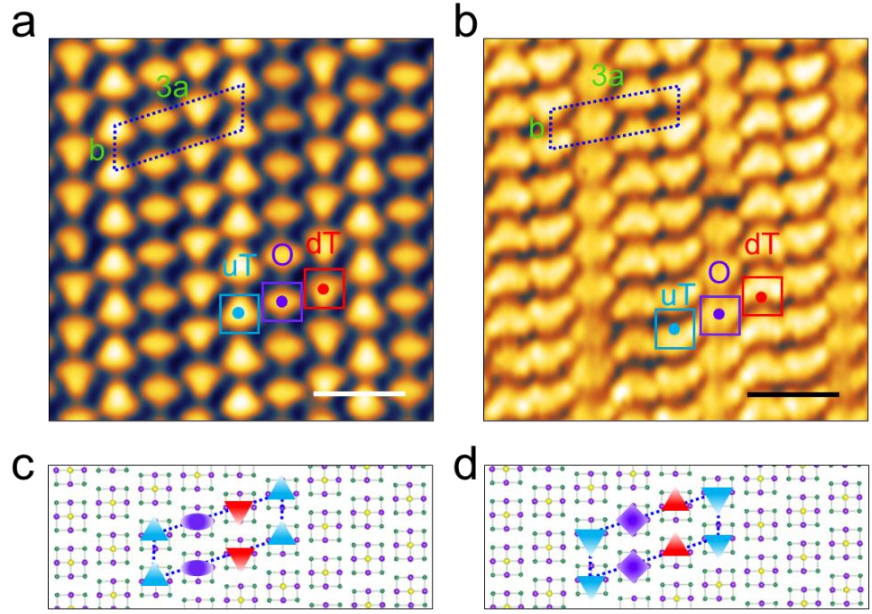

**FIG. S10:** Bias polarity-specific STM images. (a, b) Bias-dependent STM images of empty (a) and unoccupied (b) states for the 1 × 3 supercell of triple-cube CDW. The supercell of the triple-cube CDW consists of a chain of "olive" (O), "down-triangular" (dT) and "up-triangular" (uT) shapes in (a). (c) and (d) are the schematic models of (a) and (b), respectively. (a) Scale bar, 2 nm, $V_s$ = 1.2 V, (b) scale bar, 2 nm, $V_s$ = -1.2 V.

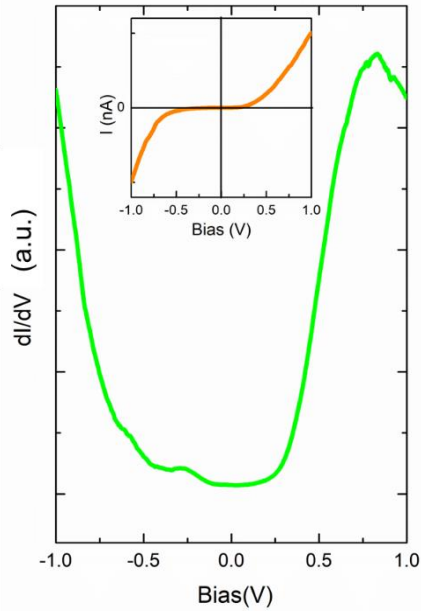

**FIG. S11:** STS spectra of ATS at ~150 K (above the charge order transition temperature). No density of states gap was observed in the STS spectra, which is markedly different from the corresponding STS spectra obtained at 9 K.

# Supplementary Note 2: tc-CDW-the bulk and/or surface property

In addition to the STM and ARPES measurements, we also performed transport and Raman measurements which primarily reflect bulk properties of the ATS crystal. Both surface and bulk sensitive measurements show the two transitions near 110~120 and 80~90 K, indicating that the transitions occur both on the surface and in the bulk. In particular, the original Fig. 2 (a,b) shows the electron transport and Raman measurement results, which clearly indicate a transition near 110~120 K, corresponding to the tc-CDW observed using STM on the surface.

As revealed in our STM images, the tc-CDW occurs in one direction of the ab-plane, which means that the CDW gap opens in a certain path of the Brillouin Zone (BZ) and the crystal remains metallic in other paths, showing the partially suppressed DOS within the tc-CDW gap in our STS spectra. This is consistent with our band structure calculations, namely the gapped Γ-X path and the metallic Γ-Y path, for the bulk (Fig. 3b, c) and monolayer (Fig. S7a-c) of ATS. The partially suppressed DOS was also found in some anisotropic bulk CDW materials such as $CeTe_3$ [9, 10]

The second-order differential conductance curve (Fig. S8b) indicates a bulk phase transition at ~ 90 K. In the same temperature window, we observed an antipolar charge order transition on the surface in our STM images (Fig. 2c-d). These results suggest that the antipolar charge order occurs both in the bulk and on the surface of the ATS crystal. This assessment is also consistent with the fact that the antipolar state is reinforced by the tc-CDW which is a property existing both in the bulk and on the surface of the ATS crystal. Although we tried to measure the dielectric properties of ATS single crystal, it is extremely challenging to obtain meaningful signals of the antipolar charge order, which is, most likely, due to the metallicity of ATS.

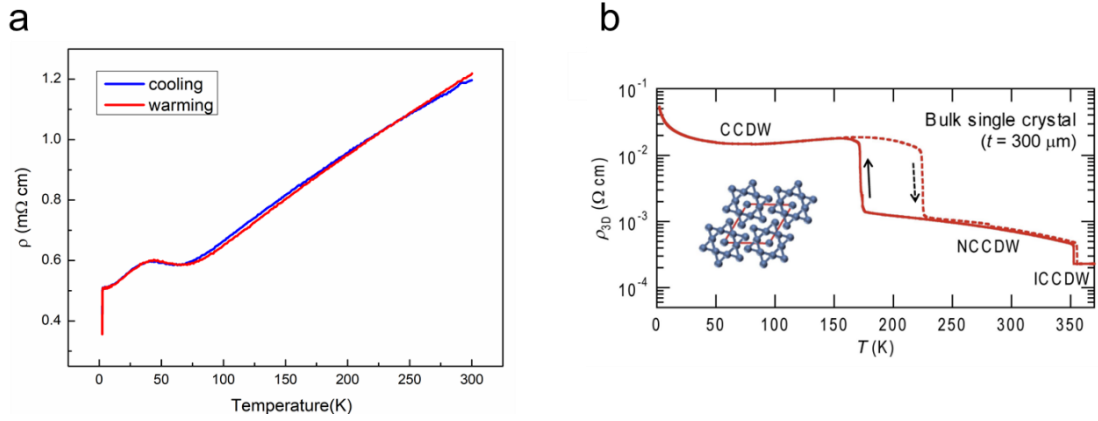

**FIG. S12:** (a) Temperature-dependent resistivity during warming (red) and cooling (blue) of ATS, in which no apparent hysteresis is observed. (b) Temperature-dependent resistivity during warming and cooling of $TaS_2$, in which temperature hysteresis can be clearly observed in the 150–250 K temperature range [11]. Measurements during warming and cooling cycles were carried out in our variable-temperature STM, electron transport and Raman spectroscopy measurements. Typical temperature-dependent resistivity curves acquired during the warming and cooling cycles were plotted in (a), showing a metallic behavior and no appreciable hysteresis. As a comparison, the curves of 1T-$TaS_2$ are shown in (b) [9], in which a hysteresis is clearly identified together with an insulator behavior. No obvious temperature hysteresis was found in the process of warming and cooling of STM and Raman measurements, suggesting the electronic nature of the two transitions. All these measurements are reproducible and both transitions are reversible.

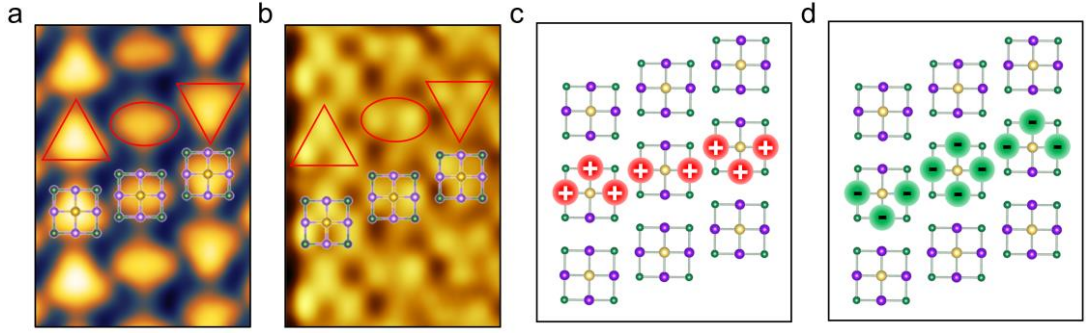

**FIG. S13:** STM topography and atomic resolution images of ATS at the triple-cube CDW. (a) A regular STM topography image of the triple-cube CDW. The (uT-O-dT) cubes show the specific up-triangular (uT), olive (O) and down-triangular (dT) shapes, respectively. (b) The atomic resolution image of the triple-cube CDW. The atomic structure of the cubes (top layer) was overlaid on the (uT-O-dT) cubes in (a) and (b). (c, d) The schematic structural models of the (uT-O-dT) cubes (top layer) for the triple charge order states obtained at positive and negative bias. Scanning parameters: (a) 2.5 × 3.5 nm, $V_s$ = +1.4 V, (b) 2.5 × 3.5 nm, $V_s$ = +2.3 V. The atomic-resolution images of ATS show that the difference in Te atoms along the *b*-axis of top layer is the reason for the polarized electronic states. At the positive bias, the up-triangular (uT), olive (O) and down-triangular (dT) shapes are mainly contributed by the 'up-three' (uT), 'middle-two' (O) and 'down-three' (dT) Te atoms of the top layer, which are highlighted by the three red circles in (c). At the negative bias, the down-triangular (uT), rhombus-shaped (O) and up-triangular (dT) shapes are mainly contributions of the "down-three" (uT), "all-four" (O) and "up-three" (dT) Te atoms of the top layer, which are highlighted by the three green circles in (d).

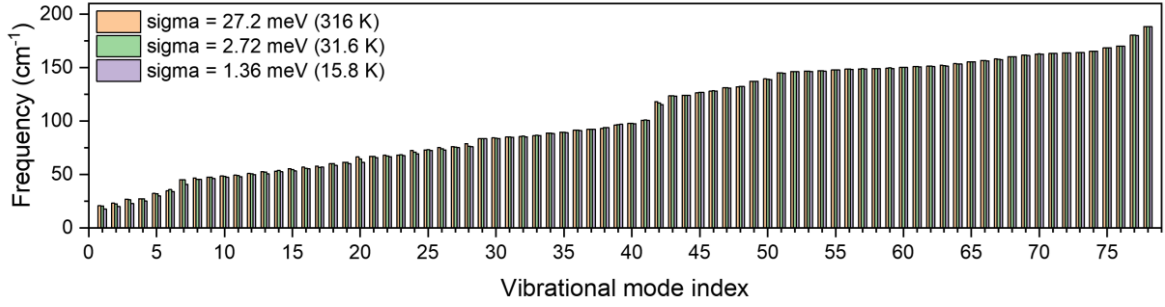

**FIG. S14:** Predicted phonon frequency of bulk ATS at $q_{tc\text{-}CDW}$ = (1/3,0,0). Calculations are performed using different smearing factors, representing low and high electronic temperatures, respectively. Softening of phonon modes are invisible among all the 78 phonon modes with the decreasing of electronic temperature. To confirm the electronic origin the tc-CDW, we performed vibrational frequency calculations of bulk ATS at $q_{tc\text{-}CDW}$ = (1/3, 0, 0) in the reciprocal space, corresponding to the three-cube periodicity in the real space, to examine whether Kohn anomaly appears. Values of 27.2 meV, 2.72 meV and 1.36 meV were used for the smearing energy, which correspond to electronic temperatures of 316 K, 31.6 K, and 15.8 K, respectively. No significant softening was found for all 78 phonon modes from 316 to 15.8 K. The lower temperature limit is far below the tc-CDW transition temperature (120 K), confirming that the electron-electron interaction is the predominant cause of the tc-CDW.

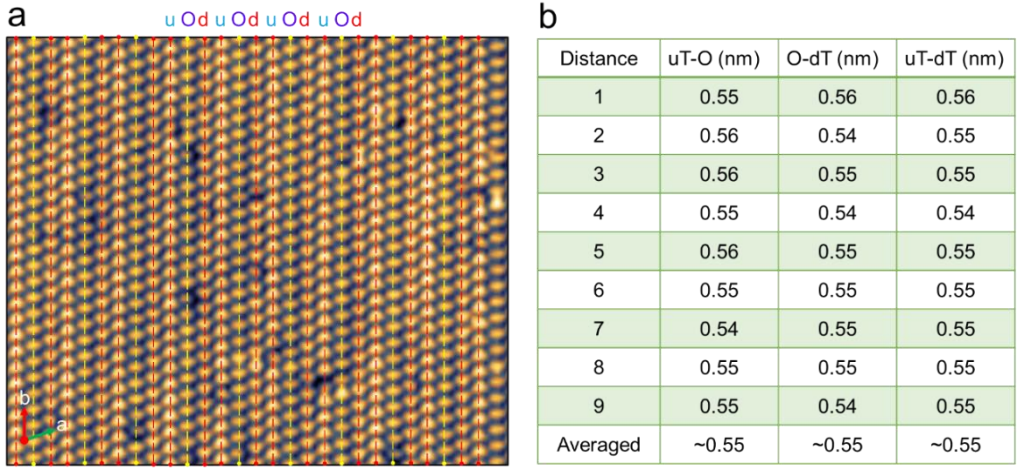

**FIG. S15:** (a) Large-scale STM topography at 9 K. (b) Statistics of the distance among ATS chains. We compare the inter-chain distances among uT-O, O-dT and uT-dT chains in the unoccupied-state STM image. We plotted a series of parallel lines passing through the center of the ATS cubes along the *b*-axis, as shown in (a). The observed mirror symmetry of each chain with respect to the *b*-axis in the image enables accurate determination of the chain (line) position and thus the inter-chain (line) distances. As listed in (b), all these inter-chain distances are ~5.5 Å with an uncertainty of ± 0.1 Å, indicating nearly identical inter-chain distances in our obtained STM images. These results confirm that no apparent cube displacements are accompanied along the tc-CDW direction.

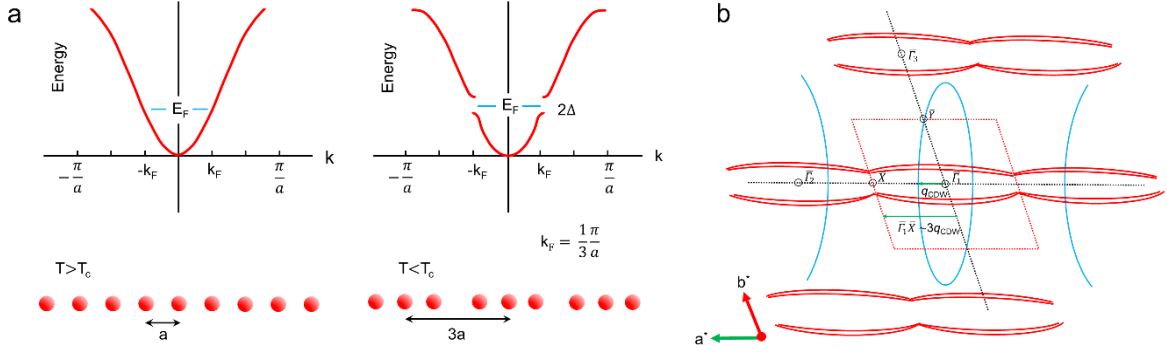

**FIG. S16:** Schematic of 1D charge density wave (CDW) mechanism in reciprocal space. (a) The schematic of a triple-cube CDW transition mechanism for 1D metal CDW with $k_F = 1/3$ ($\pi/a$). (b) A sketch of the Fermi surface (according to the ARPES results in Fig. 3e) and formation mechanism of the triple-cube CDW in ATS. The red parallelogram represents the reciprocal lattice of the *ab*-plane. The Fermi surface consists of a pair of flat sheets (red) parallel to $\overline{\Gamma}_2 - \overline{X} - \overline{\Gamma}_1$ direction and an ellipse (blue) perpendicular to $\overline{\Gamma}_2 - \overline{X} - \overline{\Gamma}_1$. The half axis of the ellipse (along the $a^*$ axis) is identified as $q_{CDW}$ (the short green arrow), which is approximately 1/3 of $\overline{\Gamma} - \overline{X}$ (the long green arrow). The $q_{CDW}$ is considered as the nesting vector for the formation of the triple-cube CDW in ATS.

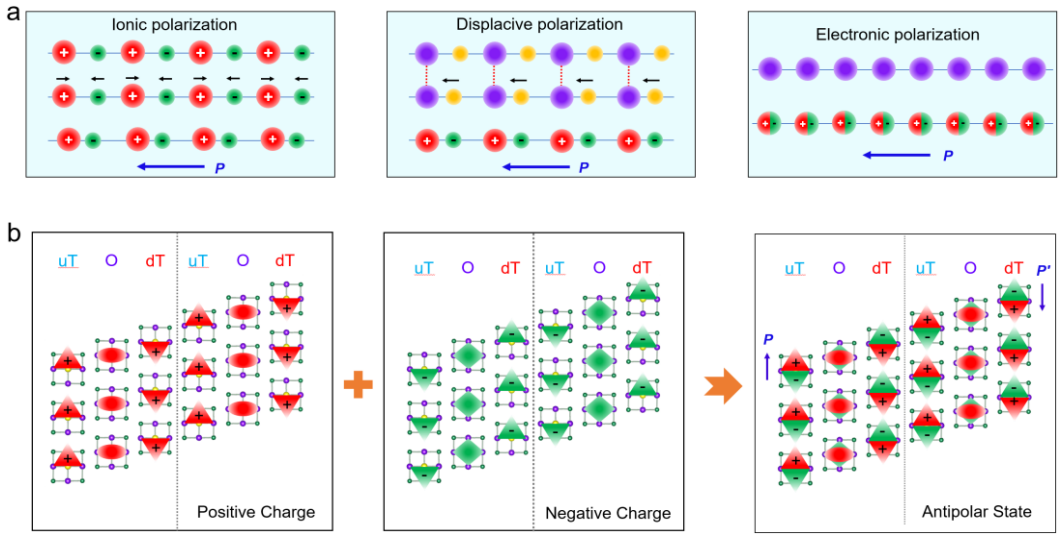

**FIG. S17:** Schematic of dipole formation mechanism and polar states. (a) Ionic polarization: the relative displacements of the positively charged and negatively charged ions. Displacive polarization: the previous neutral atoms undergo a neutral-ionic transition due to their relative displacements. Electronic polarization: the polarization (charge separation) is due to the ordering of electrons. (b) Schematic of electronic states in ATS due to the electronic polarization of super-atoms (cubes) along the *b*-axis and tc-CDW state along the *a*-axis in the low-symmetric *P*1 lattice.

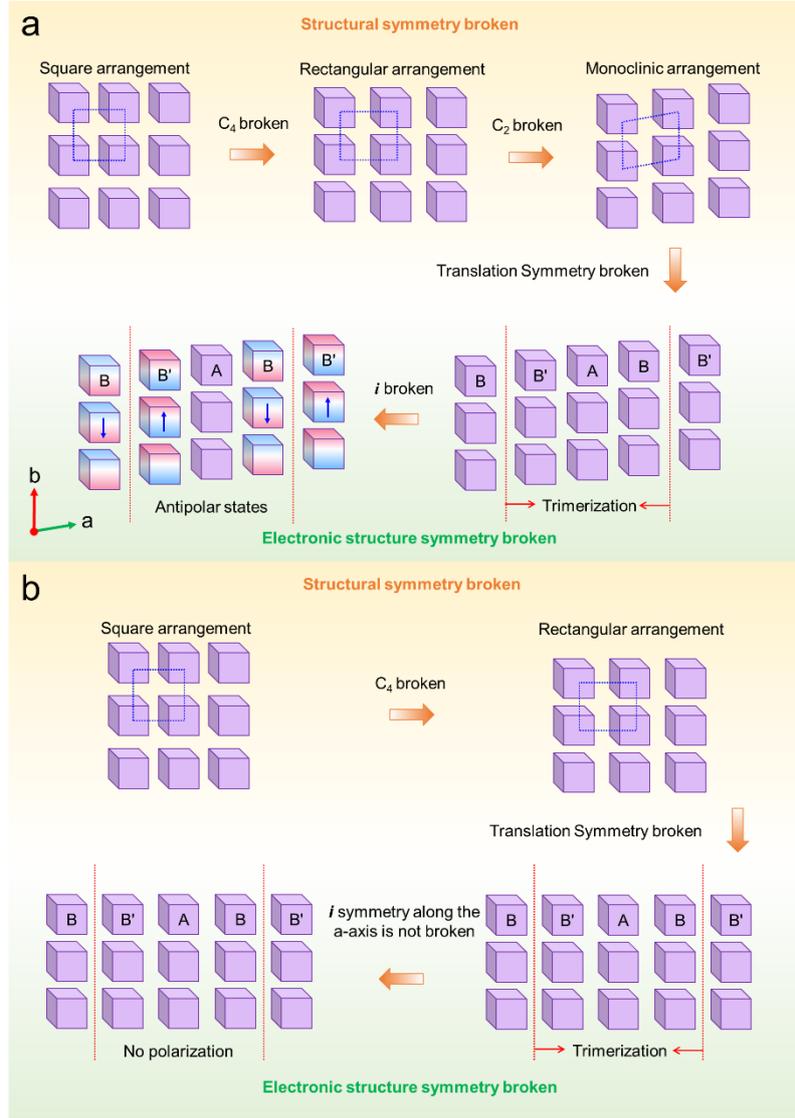

**FIG. S18:** Schematic symmetry-reducing process in the geometric and electronic structure of the layered ATS super-atomic crystal. (a) The building block is the high-symmetric cube of ATS, which could form a layered geometric structure in a square arrangement. The square arrangement of cubes is reduced to a rectangular arrangement and then monoclinic arrangement ($P1$ symmetry) by the sequential breaking of $C_4$ and $C_2$ symmetry to minimize the directive inter-cube bonding/interaction energy. The translation symmetry of monoclinic arrangement ($P1$ symmetry) is further broken to minimize the electronic energy via the formation of triple-cube CDW (along the $a$-axis). The inversion symmetry of B (and B′) cubes is broken due to the trimerization of cube along the $b$-axis, then the electronic energy is further decreased via the polarization of B (and B′) cubes and the formation of polarization states in the low-symmetric geometric structure. (b) However, the inversion symmetry of B (and B′) cubes along the $b$-axis is preserved for the rectangular arrangement ($C_2$ symmetry), which indicates that the low-symmetric geometry structure (monoclinic arrangement) is necessary for forming the polarization states in ATS.